\input phyzzx
\def\tg{\widetilde{g}}
\def\tE{\widetilde{\cal E}}
\REF\COL{%
See, for example, S. Coleman,
in {\sl The Why of Subnuclear Physics}, A. Zichichi ed.,
(Plenum, New York, 1979).}
\REF\STO{%
M. Stone,
{\sl Phys. Rev.} {\bf D14} (1976) 3568;
{\sl Phys. Lett.} {\bf 67B} (1977) 186.\nextline
S. Coleman,
{\sl Phys. Rev.} {\bf D15} (1977) 2929.\nextline
C. Callan and S. Coleman,
{\it ibid.} {\bf D16} (1977) 1762.}
\REF\NAK{%
T. Nakamura, A. Ottewill, and S. Takagi,
``The Bounce and Quasi-Stationary State Methods in the Theory of
Quantum Decay,'' Tohoku University report (1996).}
\REF\REVIEW{%
For a review, see,
{\sl Large-Order Behavior of Perturbation Theory},
J. C. Le Guillou and J. Zinn-Justin eds.,
(North-Holland, Amsterdam, 1990).}
\REF\KLE{%
H. Kleinert,
{\sl Phys. Lett.} {\bf B300} (1993) 261;
{\it ibid.} {\bf A190} (1994) 131.\nextline
R. Karrlein and H. Kleinert,
{\it ibid.} {\bf A187} (1994) 133.\nextline
H. Kleinert and I. Mustapic,
{\sl Int. J. Mod. Phys.} {\bf A11} (1996) 4383,
and references therein.}
\REF\OURS{%
H. Suzuki and H. Yasuta,
{\sl Phys. Lett.} {\bf B400} (1997) 341.}
\REF\KLA{%
J. R. Klauder,
{\sl Acta Phys. Austriaca Supple.} {\bf 11} (1973) 34.\nextline
G. Parisi,
{\sl Phys. Lett.} {\bf 67B} (1977) 167.\nextline
C. Itzykson, G. Parisi, and J. B. Zuber,
{\sl Phys. Rev. Lett.} {\bf 38} (1977) 306.\nextline
S. Coleman, V. Glaser, and A. Martin,
{\sl Comm. Math. Phys.} {\bf 58} (1978) 211.}
\REF\BRE{%
E. Br\'ezin and G. Parisi,
{\sl J. Stat. Phys.} {\bf 19} (1978) 269.}
\REF\ECK{%
J. P. Eckman, J. Magnen and R. S\'en\'eor,
{\sl Commun. Math. Phys.} {\bf 39} (1975) 251.\nextline
J. S. Feldman and K. Osterwalder,
{\sl Ann. Phys.} (NY) {\bf 97} (1976) 80.\nextline
J. Magnen and R. S\'en\'eor,
{\sl Commun. Math. Phys.} {\bf 56} (1977) 237.}
\REF\LOE{%
J. J. Loeffel,
Centre d'Etudes Nucl\'eaires de Sacley Report DPh-T/76-20 (1976),
reprinted in~[\REVIEW].}
\REF\BAK{%
G. A. Baker, B. G. Nickel, M. S. Green, and D. I. Meiron,
{\sl Phys. Rev. Lett.} {\bf 36} (1976) 1351.\nextline
G. A. Baker, B. G. Nickel, and D. I. Meiron,
{\sl Phys. Rev.} {\bf B17} (1978) 1365.\nextline
J. C. Le. Guillou and J. Zinn-Justin,
{\sl Phys. Rev. Lett.} {\bf 39} (1977) 95;
{\sl Phys. Rev.} {\bf B21} (1980) 3976.}
\REF\BEN{%
C. M. Bender and T. T. Wu,
{\sl Phys. Rev.} {\bf 184} (1969) 1231.}
\REF\ZIN{%
J. Zinn-Justin,
{\sl J. Math. Phys.} {\bf 22} (1981) 511, and references therein.}
\REF\BOG{%
E. B. Bogomolny,
{\sl Phys. Lett.} {\bf 91B} (1980) 431.\nextline
J. Zinn-Justin,
{\sl Nucl. Phys.} {\bf B192} (1981) 125;
{\it ibid.} {\bf B218} (1983) 333;
{\sl J. Math. Phys.} {\bf 25} (1984) 549.}
\REF\NIC{%
B. G. Nickel,
{\sl J. Math. Phys.} {\bf 19} (1978) 542.}
\REF\LAU{%
B. Lautrup,
{\sl Phys. Lett.} {\bf B69} (1977) 109.\nextline
G. 't Hooft,
in {\sl The Why of Subnuclear Physics}, A. Zichichi ed.,
(Plenum, New York, 1979).\nextline
P. Olesen,
{\sl Phys. Lett.} {\bf B73} (1978) 327.\nextline
G. Parisi,
{\it ibid.} {\bf B76} (1978) 65.}
\REF\BER{%
C. Bervillier, J. M. Drouffe, J. Zinn-Justin, and C. Godr\`eche,
{\sl Phys. Rev.} {\bf D17} (1978) 2144.}
\REF\YASU{%
H. Yasuta, in preparation.}
\FIG\figone{%
Ratio of an imaginary part of the quasi-ground state energy to the
leading bounce result in $D=1$ and~$N=1$ case. The solid line is the
exact numerical value. Eq.~\sixteen\ with $P=4$ (circle), $P=5$
(filled square), and $P=15$ (filled circle), is plotted. The broken
line is the leading bounce result with the two loop
correction~\addthree.}
\FIG\figtwo{%
Same as Fig.~\figone, but for~$N=2$. Eq.~\sixteen\ with~$P=30$
(square) is also plotted for a comparison.}
\FIG\figthree{%
Same as Fig.~\figone, but with the first two perturbation
coefficients are set by hand to zero, $c_0=c_1=0$.}
\FIG\figfour{%
Vacuum bubble diagrams with the counter terms~\eighteen\ to five loop
orders.}
\FIG\figfive{%
Ratio of the imaginary part of the vacuum energy density computed
by~\sixteen\ to the leading bounce result in $D=2$ and~$N=1$ case.
$P=2$ (circle), $P=3$ (filled square), and $P=4$ (filled circle),
are plotted.}
\pubnum={%
IU-MSTP/20; hep-th/9704105}
\date={April 1997}
\titlepage
\title{%
Quantum Bubble Nucleation beyond WKB: Resummation of Vacuum Bubble
Diagrams}
\author{%
Hiroshi Suzuki\foot{%
electric mail: hsuzuki@mito.ipc.ibaraki.ac.jp}
and
Hirofumi Yasuta\foot{%
electric mail: yasuta@mito.ipc.ibaraki.ac.jp}}
\address{%
Department of Physics, Ibaraki University, Mito 310, Japan}
\abstract
On the basis of Borel resummation, we propose a systematical
improvement of bounce calculus of quantum bubble nucleation rate.
We study a metastable super-renormalizable field theory,
$D$~dimensional $O(N)$~symmetric $\phi^4$~model ($D<4$) with an
attractive interaction. The validity of our proposal is tested
in~$D=1$ (quantum mechanics) by using the perturbation series of
ground state energy to high orders. We also present a result
in~$D=2$, based on an explicit calculation of vacuum bubble diagrams
to five loop orders.
\endpage
\chapter{Introduction}
In this paper, we propose a new approach for the tunneling phenomenon
in a metastable super-renormalizable field theory, with aiming at a
systematical improvement of the bounce (or instanton)
calculus~[\COL,\STO]. We shall study a $D-1$~dimensional system
($D<4$) whose Hamiltonian is defined by
$$
   H=\int d^{D-1}x\,\left[
   {c^2\over2}\pi^2+{1\over2}(\partial_x\phi)^2
   +{1\over2}m^2\phi^2-{1\over4!}g(\phi^2)^2\right],\quad g>0
\eqn\one
$$
where $\phi$~is an $N$~component real scalar field
($\phi^2\equiv\phi\cdot\phi$) and $\pi$~is its conjugate momentum.
The Hamiltonian~\one\ may be regarded as a Ginsburg-Landau like
effective theory in which the classical time evolution of order
parameter~$\phi$ is determined by the first term. Our present
approach therefore might become relevant for tunneling problems in
condensed matter physics.

We study an imaginary part of the vacuum energy density. Since the
potential energy in~\one\ is not bounded from below, the quantum
tunneling makes the naive ground state (vacuum) $\phi=0$ metastable.
Therefore, physical quantities have to be defined by an
analytic continuation from a negative~$g$. In particular, the
continuation produces an imaginary part of the vacuum energy
density~${\rm Im}\,{\cal E}$, which can be regarded as a total decay
width per unit volume of the quasi-vacuum, i.e.,
${\mit\Gamma}=-2\,{\rm Im}\,{\cal E}/\hbar$\rlap.\foot{%
One can find a detailed account on this point in~[\NAK].}

The standard method for calculating such a tunneling amplitude in
field theory is the bounce calculus~[\COL,\STO]. The existence of
the bounce solution, an extreme of the Euclidean action with one
negative Gaussian eigenvalue, signifies a decay of the vacuum due
to quantum tunneling~[\COL,\STO]. The leading bounce approximation
furthermore gives a quantitatively reliable estimation of the
tunneling amplitude in the weak coupling region~$g\ll1$.

The systematical evaluation of higher order corrections to the
leading bounce result, however, is difficult. One has to consider
``interactions'' among bounces and an integration over the
quasi-collective coordinates. Simultaneously, perturbative corrections
around bounces have to be taken into account. One also has to resolve
the ``mixing'' of those two effects. Even if this difficult task
could be done, such an ``instanton expansion'' is likely to be an
asymptotic expansion, thus we do not expect the quantitative validity
in the strong coupling region.

We tackle this issue, i.e., a {\it systematical quantitative\/}
improvement of bounce calculus, from a completely different viewpoint.
Namely we utilize an information of the {\it conventional\/}
perturbation series around the {\it naive\/} vacuum to evaluate the
tunneling rate. This should sound strange because usually the quantum
tunneling is regarded as a non-perturbative phenomenon. However,
extensive studies on the large order behavior of the perturbation
series~[\REVIEW] have revealed an intrinsic connection between the
quantum tunneling and the nature of perturbation series. This
connection is the backbone of our approach. Technically, we utilize
the Borel resummation method~[\REVIEW] for extracting the tunneling
rate from the perturbation series: Singularities of the Borel
transform are expected to reproduce an imaginary part of the vacuum
energy density.

In this approach based on the conventional perturbation expansion,
every steps of calculation (such as the renormalization) are
well-understood and, in principle, the order of approximation can be
systematically increased. To our knowledge, this kind of approach to
tunneling phenomena in quantum mechanics was initiated in~[\KLE] on
the basis of another kind of resummation method. See~[\OURS] and
references therein. Our present proposal can be regarded as the
natural generalization of~[\OURS] to quantum field theory.

We will take the ``natural unit'' $\hbar=c=1$ in what follows.

\chapter{Bounce Calculus}
Let us first recapitulate the result of bounce calculus~[\COL,\STO]
of the tunneling amplitude. It gives an important information on the
nature of a Borel singularity and, simultaneously, provides the
``standard'' with which our results can be compared.

The Lorentzian action corresponding to Hamiltonian~\one\ is given by
$$
   S[\phi]=
   \int d^Dx\,\left[{1\over2}\partial_\mu\phi\partial^\mu\phi
   -{1\over2}m^2\phi^2+{1\over4!}g(\phi^2)^2\right]\equiv iS_E[\phi],
\eqn\two
$$
where we have defined the Euclidean action~$S_E[\phi]$ (the time
coordinate of Euclidean spacetime is defined by $x^D=it$). We also
assume an appropriate counter term~$S_{\rm count.}[\phi]$ to remove
the ultraviolet (UV) divergences:
$$
   S_{\rm count.}[\phi]
   =\int d^Dx\,\left[
   {N\over2}\int{d^Dk\over i(2\pi)^D}\,\ln(m^2-k^2-i\epsilon)
   +\cdots\right].
\eqn\three
$$
The structure of abbreviated terms depends on the spacetime
dimension~$D$. See Eq.~(5.1) as the example for~$D=2$.

We repeat the procedure in~[\STO] to yield the
imaginary part of the vacuum energy density for~$g\ll1$,
$$
   \left[{\rm Im}\,\tE(g)\right]_{\rm bounce}
   =-A_NC_{D,N}\left({S_0\over2\pi\tg}\right)^{(D+N-1)/2}
   e^{-S_0/\tg},
\eqn\four
$$
where dimensionless combinations $\tE\equiv{\cal E}/m^D$
and~$\tg\equiv g/m^{4-D}$ have been introduced.

We briefly explain how the various factors in~\four\ emerge. One
expands the Euclidean functional integral around the spherically
symmetric bounce $\phi_c(r)$, which is a solution of the Euclidean
equation of motion,
$-\Delta\phi_c(r)+m^2\phi_c(r)-g\phi_c(r)^3/3!=0$. The spherically
symmetric bounce has the least action~[\KLA] whose value is
numerically given by ($S_E[\phi_c]\equiv S_0/\tg$)\foot{%
The numbers, $I_4$, $I_6$, $\overline D_R(1)$ and
$\overline D_R(1/3)$ are defined in~[\BRE]: Eq.~(58) of~[\BRE] should
be read as ${\Delta_T}^\bot=(4-D)\overline D(1/3)/4$ for a
general~$D$.}
$$
   S_0={3\over2}I_4=\cases{8&for $D=1$,\cr
                           35.10269&for $D=2$,\cr
                           113.38351&for $D=3$.\cr}
\eqn\five
$$
Gaussian integrations around the bounce except the zero modes give
rise to the coefficient~$C_{D,N}$
$$
\eqalign{
   C_{D,N}&= e^{iS_{\rm count.}[\phi_c]}m^{-(D+N-1)}
   (4-D)^{(N-1)/2}
\cr
   &\quad\times
   \left|{\det}'(-\Delta+m^2-g\phi_c^2/2)\right|^{-1/2}
   \left[{\det}'(-\Delta+m^2-g\phi_c^2/3!)\right]^{-(N-1)/2}.
\cr
}
\eqn\six
$$
The determinant can be evaluated analytically in~$D=1$ and numerically
in $D=2$ and~$D=3$~[\BRE]:
$$
\eqalign{
   C_{D,N}&=
   \left[\overline D_R(1)\left(DI_4/4\over I_6-I_4\right)^D
     \right]^{-1/2}
     \left[\overline D_R(1/3)/4\right]^{-(N-1)/2}
\cr
   &=\cases{2\sqrt{3}\times(2\sqrt{3})^{N-1}&for $D=1$,\cr
            0.3503\times(1.652)^{N-1}&for $D=2$,\cr
            10.189\times(1.6569)^{N-1}&for $D=3$.\cr}
\cr
}
\eqn\seven
$$
Since the quadratic operator $-\Delta+m^2-g\phi_c^2/2$ has one
negative eigenvalue~[\STO,\KLA,\BRE], the square root of the
eigenvalue produces a factor~$\pm i$. A physical requirement that we
are concerned with a decaying process specified the branch in~\four.
The Jacobian from a zero mode to a collective coordinate
is~$\sqrt{S_0/(2\pi\tg)}$. Because there are $D+N-1$~collective
coordinates (the spacetime position and the direction of bounce in
$N$~dimensional internal space), the power in~\four\ is resulted. The
integration over the bounce direction in the internal space gives a
factor~$A_N$, the half area of $N-1$~dimensional unit sphere,
$A_N\equiv\pi^{N/2}/\Gamma(N/2)$. Finally the integration over the
position of bounce produces a factor, the system volume times the
time period, which is divided to give the energy density: This
completes our quick review of the bounce result~\four.

In our approach, we do not need the explicit value of the determinant
factor~\seven. It will be used merely for a comparison. On the other
hand, the value of the bounce action~\five\ and the number of
collective coordinates are important because they tell the position
and the strength of the nearest Borel singularity to the origin of
Borel plane\rlap.\foot{%
However, if one has perturbative coefficients to {\it sufficiently\/}
higher orders, even those information from bounce calculus may be
deduced from the perturbation series. See~[\OURS]}
They can be easily obtained: One only has to solve a one dimensional
differential equation to find the bounce action, and a symmetrical
consideration fixes the number of corrective coordinates.

\chapter{Resummation of Vacuum Bubbles}
The leading bounce result~\four\ is reliable for the weak
coupling region $\tg\ll1$. In this section, we present our
proposal which is expected to work even in the strong coupling
region.

We start with the {\it conventional\/} perturbative expansion of the
vacuum energy density, namely a sum of the vacuum bubble diagrams,
$$
   \tE(g)\sim\sum_{n=0}^\infty c_n\tg^n,
\eqn\nine
$$
where we have assumed an appropriate renormalization which
makes~$c_n$s finite. For super-renormalizable cases~$D<4$, only first
several $c_n$s are UV divergent.

{}From the perturbation series~\nine, we construct the Borel (more
precisely Borel-Leroy) transform:
$$
   B(z)\equiv\sum_{n=0}^\infty{c_n\over\Gamma(n+(D+N)/2)}z^n.
\eqn\ten
$$
The argument of gamma function in the denominator has been chosen
so that a singularity of the Borel transform nearest to the origin
becomes a square root branch point (see~(3.5)).

When the coupling constant $g$~is {\it negative\/} in \one\ and~\two,
namely with a bounded potential, the proof of the Borel summability
of Green's functions~[\ECK] may be generalized to the vacuum energy
density. With our definition~\ten, the Borel summability is expressed
as
$$
   \tE(g)={1\over(-\tg)^{(D+N)/2}}\int_0^\infty dz\,e^{z/\tg}
   \,z^{(D+N)/2-1}B(-z).
\eqn\eleven
$$
Now, we assume that the Borel transform~$B(z)$ is analytic on the
complex $z$~plane and possible singularities of~$B(z)$ exist only on
the positive real axis. Precisely the same assumption on the Borel
transform of Green's functions has been made in the perturbation
approach to the critical phenomena~[\BAK,\REVIEW]. We also assume the
behavior of~$B(z)$ at~$z=\infty$ is moderate enough and the
convergence of the integral~\eleven\ is determined by the exponential
factor.

Under these assumptions, we may rotate the integration contour
in~\eleven\ as $z\to e^{\mp\pi i}z$ as it goes along the upper
(lower) side of the positive real axis. This operation gives rise to
the analytic continuation for~$g>0$,
$$
   \tE(g)={1\over\tg^{(D+N)/2}}\int_0^\infty dz\,e^{-z/\tg}
   \,z^{(D+N)/2-1}B(z\pm i\varepsilon),
\eqn\twelve
$$
which is our basic equation.

With~\twelve, the leading bounce result~\four\ of the imaginary
part for~$\tg\ll1$ implies that there exists a singularity of Borel
transform at~$z=S_0$~[\REVIEW]. Because of the structure of~\twelve,
it is the nearest singularity to the origin. A comparison of \four\
with~\twelve\ shows that the singularity is a square root branch
point:
$$
   B(z)=-{1\over\sqrt{\pi}}
   A_NC_{D,N}{S_0^{1/2}\over(2\pi)^{(D+N-1)/2}}(S_0-z)^{-1/2}+\cdots.
\eqn\thirteen
$$
When substituted in~\twelve, the branch cut~\thirteen\ reproduces
the imaginary part~\four\ for~$\tg\ll1$. The sign of the
singularity~\thirteen\ itself is not determined solely from the
comparison of \four\ with~\twelve\ because there are two possible
choices of the integration contour (upper side or lower side of the
positive real axis). As a consistency test, we note that the
expansion of~\thirteen\ with respect to~$z$ yields negative $c_n$s in
view of~\ten, which is nothing but the large order behavior of the
perturbation series~[\REVIEW]. As we will see below, this is a
correct property of actual perturbation coefficients when we follow
the $-i\epsilon$~prescription. This consideration also shows that we
have to take the {\it upper\/} contour to obtain a {\it negative\/}
imaginary part, which is the physical one.

Basically we construct the Borel transform~\ten\ from the actual
perturbation series~$c_n$ and substitute it in~\twelve\ for
extracting the ``non-perturbative'' information. However the radius
of convergence of the series~\ten\ is finite~($=S_0$) due to the
singularity~\thirteen\ and we have to analytically continue the
series~\ten\ outside the convergence circle to perform the Borel
integration~\twelve. This is an impossible task without knowing all
the perturbative coefficients. However, as is well-known, this
difficulty can be avoided by the conformal mapping technique~[\LOE].
We thus introduce a new variable~$\lambda$ by
$$
   z=4S_0{\lambda\over(1+\lambda)^2}.
\eqn\fourteen
$$
The point is that the convergence circle of the series~\ten\ in
terms of~$\lambda$ is now a unit circle, within which the whole cut
$z$~plane is mapped. In particular, the real axis~$z>S_0$ is mapped
on the circle~$|\lambda|=1$. Therefore we may use a finite order
truncation of the series of~$\lambda$ in the Borel
integration~\twelve\ order by order. Then we may expect the sequence
converges to~\twelve\ (as was the case in quantum mechanics~[\OURS]).

In terms of~$\lambda$, the Borel transform~\ten\ is expressed as
$$
   B(z)=\sum_{k=0}^\infty d_k\lambda^k,\qquad
   d_k\equiv\sum_{n=0}^k(-1)^{k-n}
   {\Gamma(k+n)(4S_0)^n\over(k-n)!\,\Gamma(2n)\Gamma(n+(D+N)/2)}
   c_n.
\eqn\fifteen
$$
Combining \twelve, \fourteen, and \fifteen, and parametrizing the
unit circle by~$\lambda=e^{i\theta}$, we find the $P$th order
approximation of the imaginary part,
$$
\eqalign{
   &\left[{\rm Im}\,\tE(g)\right]_P
\cr
   &=\left({S_0\over\tg}\right)^{(D+N)/2}\int_0^\pi d\theta\,
   \exp\left(-{S_0\over\tg}{1\over\cos^2\theta/2}\right)
   {\sin\theta/2\over\cos^{D+N+1}\theta/2}
   \sum_{k=0}^Pd_k \sin k\theta.
\cr
}
\eqn\sixteen
$$
Note that this is {\it solely\/} expressed by the first
$P$~perturbative coefficients~$c_n$ and the value of bounce
action~$S_0$.

Our formula~\sixteen\ relates the value of vacuum bubble
diagrams~$c_n$ and an imaginary part of the vacuum energy density. In
quantum field theory, usually one never seriously compute the vacuum
bubble diagrams because one should be free to adjust the origin of a
{\it real\/} part of the energy density and thus it has no direct
physical meaning. On the other hand, the {\it imaginary\/} part of the
vacuum energy density has a definite physical meaning as the decay
width of the quasi-vacuum. Therefore the latter should not be modified
under a change of convention for the former. How can this general
consideration and the formula~\sixteen, which relates those two
quantities, be reconciled?

In fact, there is no real contradiction. For example, suppose that we
perform a finite renormalization of the vacuum energy density to
eliminate the first $Q$~perturbation coefficients. (Note that we
cannot take $Q=\infty$ because a simple sum of the perturbation
series is diverging.) Then the new Borel transform would be defined
by
$$
   B(z)_{\rm new}\equiv
   \sum_{n=Q+1}^\infty{c_n\over\Gamma(n+(D+N)/2)}z^n
   =B(z)-\sum_{n=0}^Q{c_n\over\Gamma(n+(D+N)/2)}z^n.
\eqn\seventeen
$$
Note that the last {\it finite\/} series (i.e., polynomial) does not
develop a singularity and thus cannot contribute to any imaginary
part via the Borel integral. Therefore the imaginary part, as it
should be, is invariant under such a change of the origin of the
energy density. Contrary to its peculiar looking, the
formula~\sixteen\ is workable in this way.

\chapter{Quantum Mechanics}
We can extensively test the validity of our master formula~\sixteen\
in quantum mechanics~$D=1$. In this case, perturbative coefficients
of the vacuum energy, namely the ground state energy, to very high
orders are available. The most efficient way for computing them is
the recursion formula method~[\BEN]. By generalizing it to the
$O(N)$~symmetric model, we have computed~$c_n$ to~$n=50$. The first
several coefficients are\foot{%
It is worthwhile to note that all the coefficients (except~$c_0$) are
proportional to~$N+2$; this holds in an arbitrary dimension~$D$.
This can be proven by noting that the partition function for~$N=-2$
can be expressed by a single component fermionic system with a trivial
interaction $(\overline\psi\psi)^2\equiv0$.}
$$
\eqalign{
   &c_0={N\over2},\quad
   c_1=-{N(N+2)\over 96},\quad
   c_2=-{N(N+2)(2N+5)\over4608},
\cr
   &c_3=-{N(N+2)(8N^2+43N+60)\over221184},
\cr
   &c_4=-{N(N+2)(168N^3+1437N^2+4270N+4420)\over42467328},
\cr
   &c_5=-{N(N+2)(1024N^4+12277N^3+57668N^2+126128N+108480)
          \over2038431744}.
\cr
}
\eqn\addtwo
$$

The exact complex quasi-ground state energy is also available by a
numerical diagonalization of the Hamiltonian in a (unconventional)
Hilbert space with a rotated boundary condition (see, for
example~[\OURS]). Therefore we can compare our formula~\sixteen\
and the bounce result~\four\ with the exact value of the imaginary
part.

In Fig.~\figone, we have plotted the result of~\sixteen\ with~$N=1$
and $P=4$, $P=5$, and~$P=15$, respectively\rlap.\foot{%
This result has already been reported in~[\OURS].}
The solid line is the exact value obtained by the numerical
diagonalization of the Hamiltonian. The imaginary part is normalized
by the leading bounce result~\four. The broken line, which is depicted
for a comparison, is the bounce result including two loop radiative
corrections~[\ZIN],
$$
\eqalign{
   &\left[{\rm Im}\,\tE(g)\right]_{\rm bounce\ plus\ two\ loops}
\cr
   &=-A_NC_{1,N}\left({S_0\over2\pi\tg}\right)^{N/2}
   e^{-S_0/\tg}
   \left(1-{21N^2+54N+20\over576}\tg\right).
\cr
}
\eqn\addthree
$$
We see an excellent convergence of~\sixteen\ to the exact value and,
as was announced, the proposal in fact gives rise to the improvement
of bounce calculus. The plotted range of the coupling constant was
determined by a criterion that the real part of ground state energy
(which can also be compute numerically) is lower than the potential
barrier. Therefore the plotted range may be regarded as the quantum
tunneling (not classically sliding) region. The agreement with the
exact value is better for $\tg$~{\it larger\/} but this is physically
reasonable: When $\tg$~becomes smaller, the potential barrier becomes
higher and wider and thus the tunneling phenomena is difficult to be
detected from a perturbative expansion around the potential origin.

Fig.~\figtwo\ is the same as Fig.~\figone, but for~$N=2$. In this
case, $P=15$~shows a small excess in the weak coupling region.
However we verified that \sixteen~eventually converges to the exact
value, as indicated by the plots of~$P=30$ (squares).

In both figures, the bounce calculus including two-loop
corrections (the broken line) is giving a rather nice fitting of the
exact value. Therefore one may wonder whether the imaginary part is
almost saturated by the perturbation corrections around the bounce,
or, the multi-bounce contribution is crucial in this coupling constant
region. It is thus of interest to study how large the multi-bounce
contribution is. We may estimate the two bounce contribution by
studying the interaction between bounces and the quasi-collective
coordinate (e.g., separation between two bounces) integration~[\BOG].
The interaction is attractive when they have the same orientation and
thus the two bounce contribution dominates the functional integral
when the separation is small. However, since the notion of
multi-bounce is meaningful only for a large separation, one may
define the partition function by an analytic continuation from the
negative~$g$~[\BOG]. In this way, we find for~$N=1$\foot{%
The sign of the contribution cannot be fixed by this prescription
alone.}
$$
   \left[{\rm Im}\,\tE(g)\right]_{\rm two\ bounces}
   =\pm{\pi\over2}\left[{\rm Im}\,\tE(g)\right]_{\rm bounce}^2
   =\pm{24\over\tg}\,e^{-16/\tg},
\eqn\addfour
$$
which is $42$\%~of the one bounce contribution for~$\tg=4$, the
boundary of the tunneling region. Therefore, if one pursues the
bounce calculus, the inclusion of multi-bounce contributions should
be crucial in this coupling constant region.

{}From the excellent agreements in Figs.\ 1 and~2, therefore we
conclude that our approach based on the perturbation series around the
trivial vacuum is in fact taking into account the multi-bounce
contributions. In other words, our assumption~\twelve\ contains a
statement that an imaginary part of the vacuum energy density is
completely saturated by Borel singularities of the perturbation
series around the {\it trivial\/} configuration.

As the final application of the quantum mechanical model, we may test
our assertion made in~\seventeen\ that a finite subtraction of the
perturbation series does not affect the imaginary part when $P$~is
sufficiently large. Fig.~\figthree\ is the same as Fig.~\figone\ but
the first two coefficients $c_0$ and~$c_1$ are set to zero
{\it by hand}. We see that the same kind of convergence behavior even
with this subtraction (except the intermediate oscillating behavior;
note that $P=4$ is better than $P=5$) and $P=15$~reproduces almost
the same result as~Fig.~\figone.

\chapter{Tunneling on Line}
After observing our proposal works quite well in quantum mechanics,
let us try to apply our formula~\sixteen\ to~$D=2$, that represents
a certain one dimensional system (line or wire).
For~$D=2$, we assume the following counter terms in~\three:
$$
\eqalign{
   &S_{\rm count.}[\phi]=\int d^2x
\cr
   &\times\biggl\{
   {N\over2}\int{d^2k\over i(2\pi)^2}\,\ln(m^2-k^2-i\epsilon)
    -{1\over12}(N+2)g\int{d^2k\over i(2\pi)^2}
    {1\over m^2-k^2-i\epsilon}\,\phi^2
\cr
   &\qquad+{1\over24}N(N+2)g\left[\int{d^2k\over i(2\pi)^2}
    {1\over m^2-k^2-i\epsilon}\right]^2\biggr\},
\cr
}
\eqn\eighteen
$$
where the first and third terms remove the zero-point energy to
two loop orders, and the second is the counter term for the one loop
self energy\rlap.\foot{%
Therefore, our coupling constant~$g$ is not the physical four point
scattering amplitude. The finite renormalization, which is required
to translate the result into the one in terms of the physical
coupling constant, might be performed by utilizing the resummed
$\beta$~function in~[\BAK].}
Since the present system is super-renormalizable, the counter
terms~\eighteen\ remove all the UV divergences to higher orders.

To find the lower order perturbative coefficients~$c_n$ in~\nine,
we have explicitly calculated the vacuum bubble diagrams to five loop
orders. With the counter terms~\eighteen, $c_0=c_1=0$, and there are
five diagrams to be evaluated (Fig.~\figfour).

\noindent
We have the following numbers: The diagram~(a),
$$
   c_2=-{N(N+2)\over3}\times8.833\,895\,1\times10^{-5}.
\eqn\nineteen
$$
The diagram~(b),
$$
   c_3=-{N(N+2)(N+8)\over27}\times3.012\,767\,294\times10^{-6}.
\eqn\twenty
$$
The diagrams (c), (d), and (e), respectively,
$$
\eqalign{
   c_4&=-{N(N+2)(N^2+6N+20)\over81}\times
   5.657\,478\,653\,058\times10^{-8}
\cr
   &\quad-{N(N+2)^2\over9}\times
   1.006\,825\,50\times10^{-7}
\cr
   &\quad-{N(N+2)(5N+22)\over81}\times
   2\times10^{-7}.
\cr
}
\eqn\twentyone
$$
In the above numbers, we have verified only the last digit contains
the roundoff error due to numerical integrations. In the
diagram~(e), we used the analytical expression of one loop triangle
diagram evaluated by~[\NIC]. The numerical error for the diagram~(e)
is much larger than others: However we verified that the final
result in~Fig.~\figfive\ is quite insensitive to the precise value
of~(e) by varying it within the estimated error. As the general
property, all the diagrams have an overall minus sign according to
the $-i\epsilon$~prescription. Notice
that we are computing the {\it energy\/} density, instead of the
effective action.

We could not prepare the exact value of the energy density, although
this task might be done by invoking a lattice formulation in a
complex rotated functional space. Therefore it is not clear which
range of~$\tg$ is the tunneling region in this case. As an {\it order
of magnitude\/} estimation, we may consider the effective potential
to two loops which behaves around the origin as
$$
   V_{\rm eff.}[\phi]=
   \left[{1\over2}
         -{1\over3}(N+2)\times6.18\times10^{-4}\,\tg^2\right]
   m^2\phi^2+\cdots,
\eqn\addone
$$
where the counter terms~\eighteen\ have been taken into account. To
the two loop order, the real part of the vacuum energy density 
vanishes, $c_0=c_1=0$. Therefore we may regard the value of~$\tg$
which changes the sign of curvature of the effective potential at the
origin as a measure of the tunneling region. This rough estimation
indicates $0<g\lsim\sqrt{3/(N+2)}\times30$ is the tunneling region.

In~Fig.~\figfive, we have plotted an imaginary part of the vacuum
energy density~\sixteen\ with~$P=2$, $P=3$, and $P=4$. Again it is
normalized by the bounce result~\four. Unfortunately, it seems
impossible to draw a definite conclusion from Fig.~\figfive:
We do not observe the convergence behavior and in some region even an
unphysical result, the positive imaginary part, can be found. It is
not clear whether this is due to the luck of orders of the
perturbation series, or there exists a fundamental obstruction for
our approach we did not encounter in quantum mechanics. (The nature
of the nearest Borel singularity in the present case, $D=2$ and
$N=1$, is the same as that of $N=2$ quantum mechanics, Fig.~\figtwo.)
If we nevertheless take Fig.~\figfive\ at its face value, it is
suggesting that the true tunneling amplitude is much {\it larger\/}
than the leading bounce result in the strong coupling region.
However, this would remain a speculation without having much higher
order perturbation coefficients, which will clarify the real
convergence property of our proposal.

\chapter{Conclusion}
In this paper, we have proposed a new approach to the tunneling
phenomena ``the decay of the false vacuum'' in a
super-renormalizable field theory. Our approach utilizes the
information of conventional perturbation series around the naive
vacuum~$\phi=0$. We have verified numerically that the singularities
of the Borel transform reproduce an accurate tunneling rate in
quantum mechanics ($D=1$). We have also presented a result in~$D=2$,
based on an explicit calculation of vacuum bubble diagrams to five
loop orders. Unfortunately the number of orders of the perturbation
series is not sufficient to make definite conclusions on the
convergence property and the true tunneling rate in this case. Only
the higher order calculation will answer these questions.

Although we have only presented results for $D=1$ and~$D=2$ in this
paper, the formula~\sixteen\ is waiting for the straightforward
application to $D=3=1+2$~dimensional system (tunneling on plane). We
expect a calculation of perturbative coefficients to six loops is
tractable because the analytical structure of one loop diagrams
in~$D=3$ is rather simpler than that of~$D=2$~[\NIC]. As a
consequence, definite conclusions on a convergence property of the
method and on the true tunneling rate may be drawn. We hope to come
back this problem in a near future.

Another possible test of our proposal is offered by an interesting
field theoretical model in~[\BER], for which both the bounce
calculus and the perturbative calculation to very high orders are
possible~[\YASU].

Finally we briefly comment on a possible generalization to the just
renormalizable case, $D=4=1+3$. The first trouble one encounters is
that all the vacuum bubble diagrams have an overall (i.e., not
sub-diagram) UV divergence. A simple renormalization which sets all
the coefficients~$c_n$ zero order by order, appears
meaningless\rlap.\foot{%
A similar situation occurs when one evaluates an imaginary part of the
vacuum energy density due to the UV renormalon. See the third
reference of~[\LAU].}
Therefore a natural procedure is that one first introduces an UV
cutoff while hoping the imaginary part~\sixteen\ is finite and
independent of the regularization for~$P\to\infty$. Next we are
worried about an emergence of the UV renormalon~[\LAU], another known
source of the Borel singularity. Interestingly, it does not emerge in
our present case because $\phi^4$~model with a {\it negative\/}~$g$
{\it is\/} asymptotically free (no Landau pole)\rlap.\foot{%
In {\it massless\/} theory, the infrared (IR) renormalon produces an
imaginary part of the vacuum energy density:
$[{\rm Im}\,{\cal E}(g)]_{\rm IR\ renormalon}=
   -1/(64\pi)\mu^4\exp[-4/(b_0g(\mu))]$,
where $\mu$~is the renormalization point and $b_0=N/[3(4\pi)^2]$
is the one loop coefficient of $\beta$~function. This gives rise
to a Borel singularity at $z=4/b_0$.}
Of course, since $\phi^4$ model in~$D=4$ is trivial when the UV
cutoff is removed, the relation to the ``true'' imaginary part is not
obvious. Presumably, our proposal for the four dimensional
$\phi^4$~model is meaningful only with an UV cutoff, but such an UV
cutoff is naturally provided in condensed matter physics.

The work of H.S. is supported in part by the Ministry of Education
Grant-in-Aid for Scientific Research, Nos.~08240207, 08640347,
and 08640348.

\refout
\figout
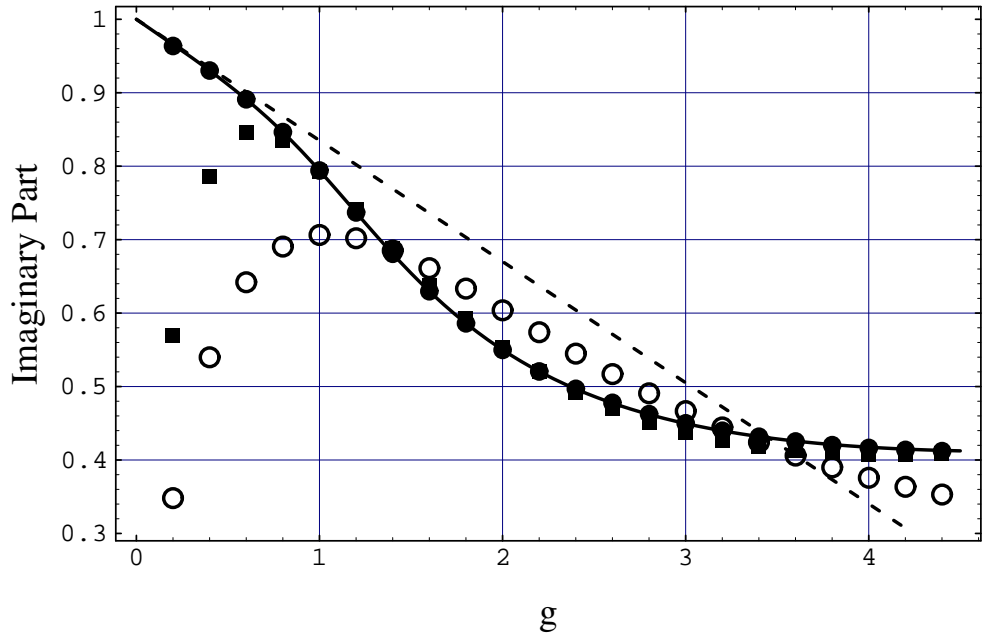
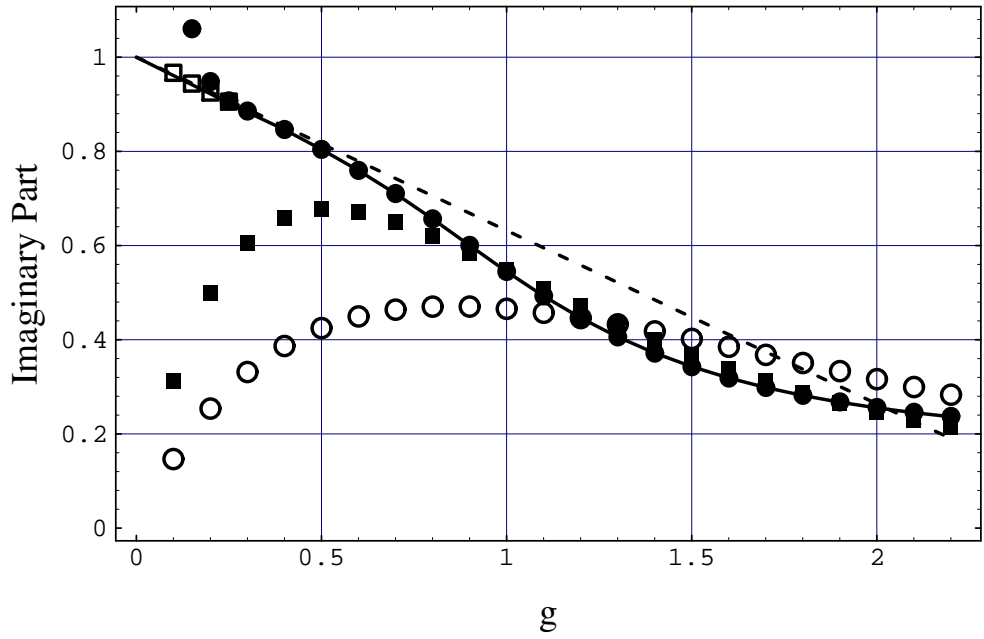
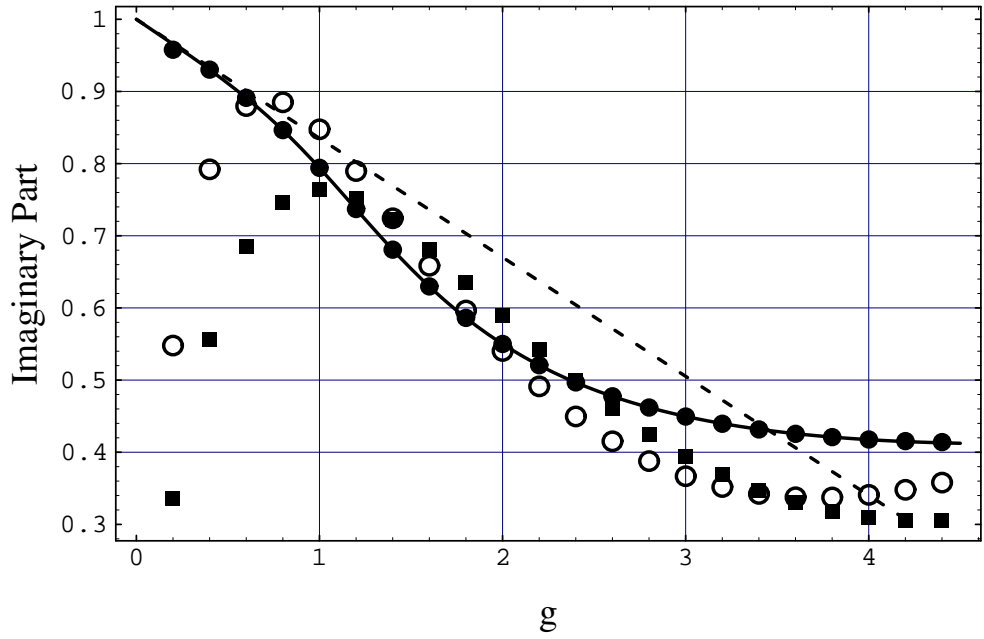
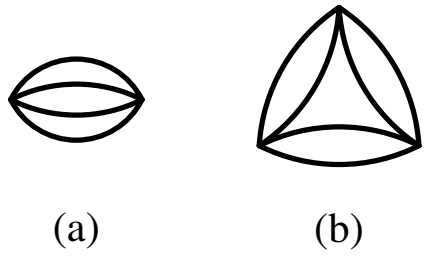
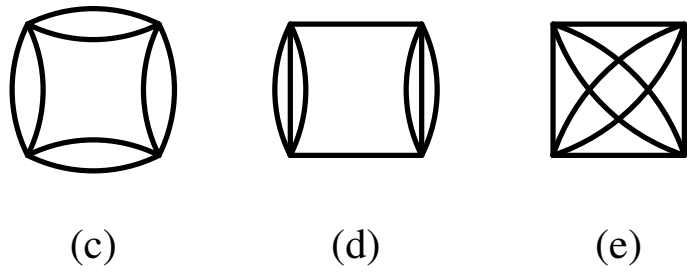
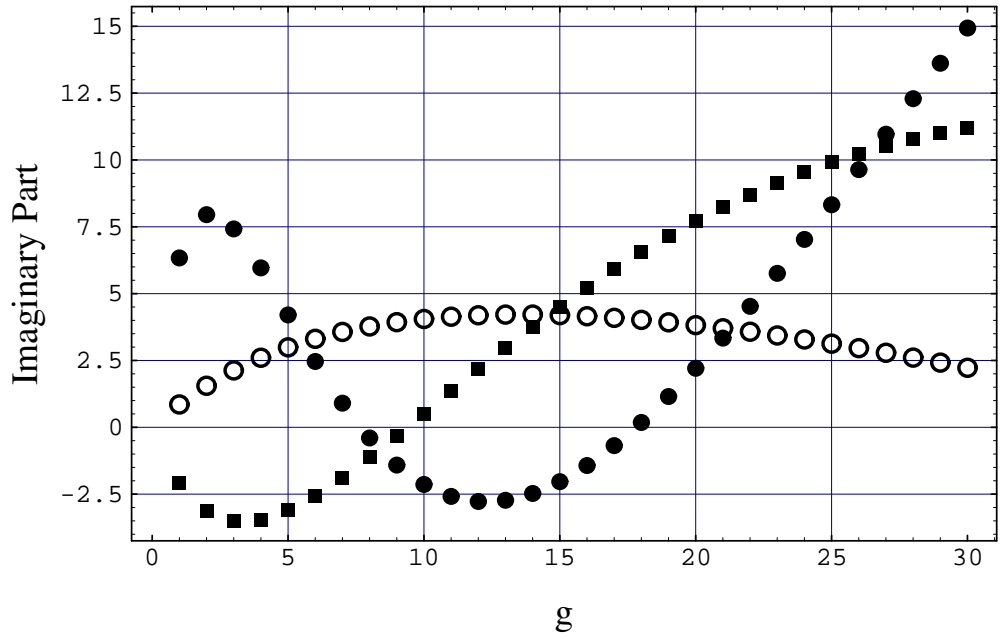
\bye